\documentclass[10pt,twocolumn]{article}

\usepackage{amsmath,amssymb,amsfonts}
\usepackage{algorithmic}
\usepackage{graphicx}
\usepackage{float}
\usepackage{multirow}
\usepackage{url}
\usepackage{adjustbox}
\usepackage{xcolor}
\usepackage{textcomp}
\usepackage[utf8]{inputenc}
\usepackage{longtable}

\usepackage{geometry}
\usepackage{fancyhdr}
\usepackage{titlesec}
\usepackage{authblk} 

\titleformat{\section}{\large\bfseries}{\thesection}{1em}{}
\titleformat{\subsection}{\normalsize\bfseries}{\thesubsection}{1em}{}

\title{\textbf{MambaRecon: MRI Reconstruction with Structured State Space Models}}
\author[]{\textbf{Yilmaz Korkmaz, Vishal M. Patel}}
\affil[]{Department of Electrical and Computer Engineering, Johns Hopkins University}
\affil[]{\{ykorkma1,vpatel36\}@jhu.edu}

\date{} 

\usepackage[pagebackref,breaklinks,colorlinks]{hyperref}

\usepackage[capitalize]{cleveref}
\crefname{section}{Sec.}{Secs.}
\Crefname{section}{Section}{Sections}
\Crefname{table}{Table}{Tables}
\crefname{table}{Tab.}{Tabs.}

\begin{document}
\def\SB#1{\textsubscript{#1}}

\maketitle     

\begin{abstract}
Magnetic Resonance Imaging (MRI) is one of the most important medical imaging modalities as it provides superior resolution of soft tissues, albeit with a notable limitation in scanning speed. The advent of deep learning has catalyzed the development of cutting-edge methods for the expedited reconstruction of MRI scans, utilizing convolutional neural networks and, more recently, vision transformers. Recently proposed structured state space models (e.g., Mamba) have gained some traction due to their efficiency and low computational requirements compared to transformer models. We propose an innovative MRI reconstruction framework that employs structured state space models at its core, aimed at amplifying both long-range contextual sensitivity and reconstruction efficacy. Comprehensive experiments on public brain MRI datasets show that our model sets new benchmarks beating state-of-the-art reconstruction baselines. Code will be available \href{https://github.com/yilmazkorkmaz1/MambaRecon}{here}.
\end{abstract}

\section{Introduction}

Magnetic Resonance Imaging (MRI) is a highly prevalent imaging technique, favored for its superior soft tissue visualization capabilities. However, its use is limited by lengthy and expensive scanning process. As a result, there is a pressing need for accelerated MRI techniques to enhance its practical application in clinical settings. Sampling less k-space points (i.e., Fourier coefficients) below the Nyquist rate can expedite the scanning but it requires to solve an ill-posed reconstruction problem \cite{lustig2007sparse}. Compressed sensing (CS) solutions have been demonstrated to be powerful in reconstruction problems 
\cite{otazo2010combination,donoho2009message,lustig2008compressed,haldar2010compressed}. However, despite their efficacy, CS methods suffer from their high sensitivity to hyper-parameters, which requires careful tuning to achieve optimal performance. Moreover, the iterative nature of most CS algorithms contributes to their computational inefficiency, making them inherently slow compared to other approaches.

Following the emergence of deep learning, iterative approaches have been increasingly supplanted by deep learning methods \cite{Wang2016, ChulYe2018, rgan, lee2018deep}. These methods predominantly leverage convolutional neural networks (CNNs) which have proven to be highly effective in various image processing tasks with data-driven approaches. Along with the pure data-driven methods, various physics-guided reconstruction models have been proposed \cite{yaman2020,MoDl,schlemper2017deep,qin2018convolutional}. In these models, the forward encoding operator, which includes the undersampling pattern and coil sensitivities in multi-coil setting, is also utilized in the model along with the under-sampled acquisitions to enhance the robustness and superior performance in the reconstruction \cite{yaman2020}. However, CNNs are inherently constrained by their limited receptive fields, which restrict their ability to capture long-range dependencies and contextual information comprehensively \cite{Zhang2019}.   

In recent years, following the success of transformers in downstream computer vision tasks, transformer-based approaches have been introduced to address the limited receptive field of CNNs in the realm of MRI reconstruction \cite{slater,guo2023reconformer,zhou2022dsformer,huang2022swin}. By incorporating transformer architectures, increased sensitivity to long-range dependencies and expanded effective receptive fields have been achieved. However, quadratic complexity of self-attention transformers with respect to input sequence length creates a computational burden \cite{keles2023computational,slater}. 

Structured state space models (SSMs) are one of the proposed solutions to address the quadratic complexity of self-attention transformers to efficiently model long sequences \cite{gu2021efficiently}. More recently, a new class of structured state space models (i.e., Mamba) has been proposed with input dependent parametrization and linearly scaled complexity with sequence length \cite{gu2023mamba}. Mamba has shown a promising direction in large language models and recently been utilized in several vision tasks including classification \cite{liu2024vmamba}, image segmentation \cite{wang2024mamba,ma2024u} and image restoration \cite{guo2024mambair}. 

Inspired from these developments, in this paper we propose a physics-guided MRI reconstruction framework utilizing structured state space models in its core with a very few number of trainable parameters. Our model benefits from the increased long-range contextual sensitivity along with the sub-quadratic complexity without omitting the underlying physical model by alternating between data-consistency and Mamba blocks. Comprehensive experiments demonstrate our model's superior performance compared with the state-of-the-art MRI reconstruction models. 

This paper makes the following  contributions:

\begin{enumerate}
    \item We propose a novel lightweight physics-guided MRI reconstruction model utilizing structured state space models as a core block.  Our model enhances the long range sensitivity of the model by significantly increasing the effective receptive field.
    \item We obtain superior performance in both complex multi-coil and magnitude single-coil public brain MRI datasets compared to the state-of-the-art reconstruction baselines.
\end{enumerate}

\section{Related Works}

\paragraph{Data-Driven Convolution Based Models.}
Convolutional neural networks (CNNs) have been predominantly used in deep MRI reconstruction to learn the underlying mapping between under- and fully-sampled acquisitions. Wang et al.\cite{Wang2016} proposed one of the first CNN-based framework for MRI reconstruction. Ye et al. \cite{ChulYe2018} proposed a CNN-based model to efficiently solve general inverse problems. Lee et al. \cite{lee2018deep} offered a deep CNN-based model with residual connections for MRI reconstruction. Hyun et al. \cite{Hyun2018} utilized a reconstruction model employing a U-shaped network. Dar et al. \cite{rgan} proposed a conditional generative adversarial network (GAN) following the success of GANs in image generation. Gungor et al. \cite{gungor2023adaptive} proposed a test time adaptation approach with a diffusion-based model. 

\paragraph{Physics-guided Convolution Based Models.}
 Aggarwal et al. \cite{MoDl} proposed an unrolled model with a CNN backbone. Schlemper et al. \cite{schlemper2017deep} utilized a deep cascaded CNN to efficiently incorporate forward operator. Eo et al. \cite{eo2018kiki} proposed a dual-domain physics-guided CNN-model. Yaman et al. \cite{yaman2020} offered an innovative solution to eliminate supervision requirement via a self-supervised approach with an unrolled CNN backbone. Yang et al. \cite{yang2018admm} proposed an unified physics-guided model trained with alternating direction method of multipliers (ADMM) algorithm. Biswas et al. \cite{Biswas2019} proposed an extra smoothness regularization with a physics-guided network. Qin et al. \cite{Conv_recur} proposed a recurrent convolutional model for dynamic MRI reconstruction. Sriram et al. \cite{Variatonal_end2end} proposed an unrolled variational model utilizing both k-space and image domains. Yiasemis et al. \cite{yiasemis2022recurrent} proposed a hybrid architecture performing optimization in k-space using an image domain refinement network.  

\paragraph{Transformer-based Models.}
Following the success of transformer-based approaches in computer vision, Korkmaz et al. \cite{slater} proposed a zero-shot learned transformer model for unsupervised reconstruction. Guo et al. \cite{guo2023reconformer} offered a phyics-guided model with a recurrent transformer backbone. Zhou et al. \cite{zhou2023dsformer} proposed a dual-domain transformer model using both frequency and image domains. Huang et al. \cite{huang2022swin} adapted the computationally efficient Swin Transformer model for MRI reconstruction. Hu et al. proposed \cite{hu2022trans} a transformer based model with a special regularization to keep high frequency details. Fabian et al. \cite{fabian2022humus} proposed an unrolled hybrid architecture employing transformers along with CNNs. Lyu et al. \cite{lyu2023region} offered a region-focused GAN with a transformer backbone. Zhao et al. \cite{zhao2023swingan} offered an adversarial model with a swin transformer backbone for MRI reconstruction. 

\paragraph{Mamba in Medical Imaging.}
More recently, a couple of Mamba-based deep learning models have been utilized in various medical imaging tasks. Yue et al. \cite{yue2024medmamba} proposed a vision Mamba model for medical image classification. Xing et al. \cite{xing2024segmamba} offered a U-shaped mamba model for 3D image segmentation. Yang et al. \cite{yang2024vivim} adapted Mamba for medical video object segmentation. Wang et al. \cite{wang2024mamba} and Ruan et al. \cite{ruan2024vm} proposed U-shaped Mamba-based models for medical image segmentation. Atli et al. \cite{atli2024i2i} proposed a Mamba-based synthesis model for medical image translation. 

Concurrently, Huang et al. 
\cite{huang2024mambamir} offered an arbitrarily masked Mamba-based model for medical image reconstruction and uncertainty estimation without taking physical model into account. Zou et al. \cite{zou2024mmr} proposed a multi-domain Mamba-based reconstruction framework, however this work differs from our approach due to usage of fully-sampled auxiliary modality along with the target reconstruction. To best our knowledge, this is the first physics-guided SSM-based model for MRI reconstruction.

\section{Background}
\subsection{Accelerated MRI Reconstruction}
MRI reconstruction from partial k-space can be defined as follows
\begin{equation}
\label{eq:sampling}
S_{p} C x = y_{p},
\end{equation}
where $S_{p}$ is the partial Fourier operator, $x$ is the target MR image,  $C$ is the coil sensitivity maps and $y_{p}$ stands for the partially sampled k-space points. Reconstruction of $x$ from $y_p$ leads to infinitely many solutions, thus solution requires regularization. Reconstruction objective with regularization can be formulated as
\begin{equation}
\widehat{x}=\underset{x}{\operatorname{argmin}}\frac{1}{2}\|y_{p}-S_{p} C x \|^{2} + \beta(x),
\end{equation}
where $\widehat{x}$ is the reconstructed MR image using the algorithm, and $\beta(x)$ is the regularization term that is used to shrink the solution space. Traditionally, total variation using gradients in the image \cite{block2007undersampled}, $\mathcal{L}_1$ norm in the Wavelet domain \cite{lustig2007sparse} or $\mathcal{L}_2$ norm of the reconstruction \cite{pruessmann1999sense} have been widely adopted as regularization terms. However, the classic approach for deep reconstruction models is to learn $\beta(x)$ during the offline training.

\begin{figure*}[h]
    \centering
    \includegraphics[width=1\textwidth]{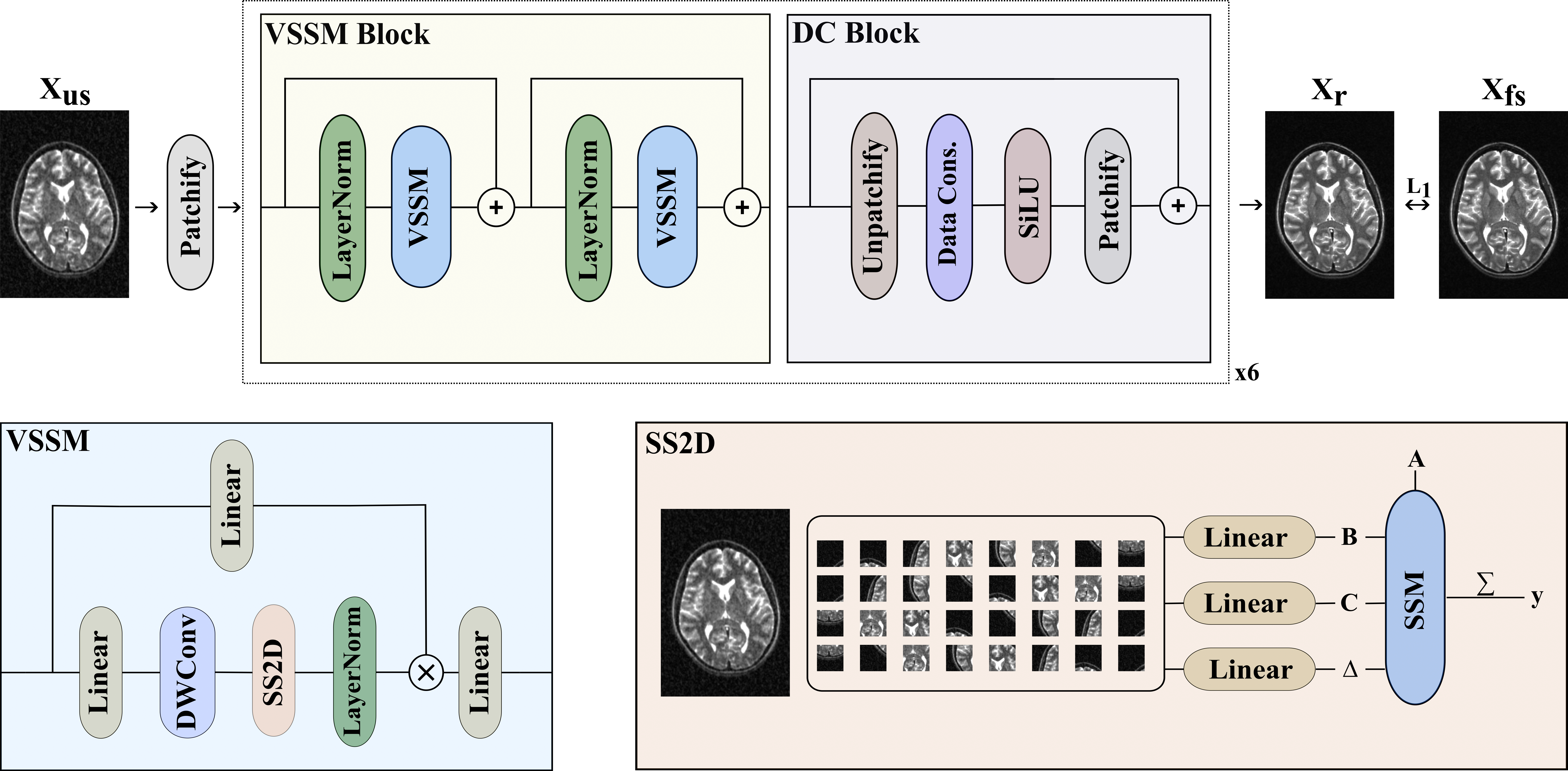}
    \caption{MambaRecon architecture is shown. $\text{X}_{\text{us}}$ corresponds to zero-filled input image, $\text{X}_{\text{fs}}$ is the fully sampled ground truth and $\text{X}_{\text{r}}$ is the reconstructed image. Minimization of $L_1$ norm of the difference between $\text{X}_{\text{fs}}$ and $\text{X}_{\text{r}}$ is utilized as the training objective. Consecutive VSSM and data-consistency blocks (shown as DC Block) are repeated 6 times. $\sum$ corresponds to summation across all unfolded vectors in the output of SSM. DWConv corresponds to the depth-wise convolution \cite{howard2017mobilenets}.}
    \label{fig:main}
\end{figure*}

\subsection{Structured State Space Models}
Structured State Space Models (SSMs) are used for 1D sequence $x(t) \in \mathbb{R}$ to sequence $y(t) \in \mathbb{R}$ transformation using an implicit latent space $h(t) \in \mathbb{R}^N$ as follows
\begin{align}
& h^{\prime}(t)=\mathbf{A} h(t)+\mathbf{B} x(t) 
\label{ssm_eq1}
\\
& y(t)=\mathbf{C} h(t),
\label{ssm_eq2}
\end{align}
where $A \in \mathbb{R}^{N \times N}$, $B \in \mathbb{R}^{N \times 1}$ and $C \in \mathbb{R}^{1 \times N}$. Discretization is needed to integrate this continuous system to the existing deep learning frameworks. Discretization is generally performed via zero-order hold (ZOH) where $\Delta$ parameter comes into space to redefine continuous parameters $A$ and $B$ in the discrete form as follows
\begin{align}
& \overline{A}=\exp (\Delta A) \\
& \overline{B}=(\Delta A)^{-1}(\exp (\Delta A)-I) \cdot \Delta B.
\end{align}
With this, we can rewrite the equations \eqref{ssm_eq1} and \eqref{ssm_eq2} as follows
\begin{align}
& h_t=\overline{A} h_{t-1}+\overline{B} x_t \\
& y_t=C h_t.
\end{align}
The selective scan mechanism proposed in Mamba makes the learnable parameters input dependent (dynamic) where $B \in \mathbb{R}^{Bs \times L \times N}$, $C \in \mathbb{R}^{Bs \times L \times N}$ and $\Delta \in \mathbb{R}^{Bs \times L \times D}$ are defined using linear projections of the input sequence $x \in \mathbb{R}^{Bs \times L \times D}$ where $Bs$ corresponds to batch size, $L$ to sequence length, $D$ to number of channels in the input and $N$ to inner state dimension.

\begin{table*}[]
\caption{Reconstruction metrics shown in the fastMRI dataset when R = 4 and 8.}
\resizebox{\textwidth}{!}{%
\begin{tabular}{|c|cccccccccccc|}
\hline
                & \multicolumn{2}{c|}{T\SB{1}-4x}                                                & \multicolumn{2}{c|}{T\SB{1}-8x}                                                & \multicolumn{2}{c|}{T\SB{2}-4x}                                                & \multicolumn{2}{c|}{T\SB{2}-8x}                                                & \multicolumn{2}{c|}{Flair-4x}                                             & \multicolumn{2}{c|}{Flair-8x}                        \\ \hline
                & \multicolumn{12}{c|}{PSNR (dB) $\vert$ SSIM (\%)}                                                                                                                                                                                                                                                                                                                                                                                                      \\ \hline
UNET            & \multicolumn{1}{c|}{30.93}          & \multicolumn{1}{c|}{86.55}          & \multicolumn{1}{c|}{27.94}          & \multicolumn{1}{c|}{80.97}          & \multicolumn{1}{c|}{28.65}          & \multicolumn{1}{c|}{83.82}          & \multicolumn{1}{c|}{26.23}          & \multicolumn{1}{c|}{78.07}          & \multicolumn{1}{c|}{28.09}          & \multicolumn{1}{c|}{83.75}          & \multicolumn{1}{c|}{25.32}          & 76.70          \\ \hline
E2E-Varnet      & \multicolumn{1}{c|}{41.30}          & \multicolumn{1}{c|}{96.76}          & \multicolumn{1}{c|}{36.11}          & \multicolumn{1}{c|}{93.78}          & \multicolumn{1}{c|}{36.64}          & \multicolumn{1}{c|}{95.82}          & \multicolumn{1}{c|}{33.19}          & \multicolumn{1}{c|}{92.86}          & \multicolumn{1}{c|}{36.91}          & \multicolumn{1}{c|}{94.60}          & \multicolumn{1}{c|}{32.33}          & 89.44          \\ \hline
RecurrentVarnet & \multicolumn{1}{c|}{43.29}          & \multicolumn{1}{c|}{97.44}          & \multicolumn{1}{c|}{38.09}          & \multicolumn{1}{c|}{95.05}          & \multicolumn{1}{c|}{38.80}          & \multicolumn{1}{c|}{96.95}          & \multicolumn{1}{c|}{34.83}          & \multicolumn{1}{c|}{94.39}          & \multicolumn{1}{c|}{38.54}          & \multicolumn{1}{c|}{95.78}          & \multicolumn{1}{c|}{34.11}          & 91.40          \\ \hline
SwinUNET        & \multicolumn{1}{c|}{35.89}          & \multicolumn{1}{c|}{93.53}          & \multicolumn{1}{c|}{33.04}          & \multicolumn{1}{c|}{91.06}          & \multicolumn{1}{c|}{32.42}          & \multicolumn{1}{c|}{91.95}          & \multicolumn{1}{c|}{30.28}          & \multicolumn{1}{c|}{88.99}          & \multicolumn{1}{c|}{31.93}          & \multicolumn{1}{c|}{90.31}          & \multicolumn{1}{c|}{29.16}          & 85.90          \\ \hline
SwinMR          & \multicolumn{1}{c|}{38.42}          & \multicolumn{1}{c|}{94.03}          & \multicolumn{1}{c|}{35.28}          & \multicolumn{1}{c|}{91.57}          & \multicolumn{1}{c|}{33.90}          & \multicolumn{1}{c|}{92.74}          & \multicolumn{1}{c|}{31.22}          & \multicolumn{1}{c|}{89.65}          & \multicolumn{1}{c|}{34.20}          & \multicolumn{1}{c|}{91.15}          & \multicolumn{1}{c|}{31.09}          & 86.39          \\ \hline
MambaRecon      & \multicolumn{1}{c|}{\textbf{43.93}} & \multicolumn{1}{c|}{\textbf{97.63}} & \multicolumn{1}{c|}{\textbf{39.08}} & \multicolumn{1}{c|}{\textbf{95.43}} & \multicolumn{1}{c|}{\textbf{39.43}} & \multicolumn{1}{c|}{\textbf{97.45}} & \multicolumn{1}{c|}{\textbf{35.47}} & \multicolumn{1}{c|}{\textbf{95.25}} & \multicolumn{1}{c|}{\textbf{39.21}} & \multicolumn{1}{c|}{\textbf{96.36}} & \multicolumn{1}{c|}{\textbf{34.84}} & \textbf{92.28} \\ \hline
\end{tabular}%
}
\label{tab:fastmri}
\end{table*}

\section{Methodology}
We adapt the Visual State Space Model (VSSM) blocks proposed in \cite{liu2024vmamba} for our physics-guided model, as shown in Figure \ref{fig:main}. Following VSSM, we simply divide the input coil-combined complex MR images into patches with size $p$ and perform unfolding in VSSM using 4 different directions (top-left to bottom-right, bottom-right to top-left, top-right to bottom-left, and bottom-left to top-right) as shown in Figure \ref{fig:main}. Unfolding in different directions is needed when employing SSMs in visual data, since the original SSMs are causal and span the input sequence in a single direction which is suitable for time-series data or language processing but contradicts the non-causal nature of images. We use $L_1$ norm of the difference between reconstructed and fully-sampled coil-combined images as the training objective, while employed AdamW optimizer is enforcing a decoupled weight decay \cite{loshchilov2017decoupled} on the network weights. Our simplified training objective can be written as follows

\begin{equation}
    \theta^* = \arg\min_{\theta} \mathbb{E}_{(x_{us}, x_{fs}) \sim p_{\text{data}}} \left[ \| x_{fs} - x_r(x_{us}, \mathcal{M}, C; \theta) \|_1 \right]
\end{equation}
where $x_{us}$ and $x_{fs}$ refers to the paired under- and fully-sampled coil-combined MR images, $x_r$ refers to the reconstructed image, $\mathcal{M}$ denotes the undersampling mask, $C$ denotes the coil sensitivities, and $\theta$ refers to the model parameters. 

To incorporate the complex MRI data, we convert complex MR images to real images using 2 separate channels for the real and imaginary components. In data-consistency blocks, we project patchified images back to the original shape with 2-channels via fully-connected layer, to convert them back to the complex form, and take centered Fourier transform. Then, we replace the generated k-space points with the acquired k-space points using the undersampling mask and the coil sensitivity maps, which are estimated using ESPIRIT \cite{uecker2014espirit} with default parameters, as shown in \cref{eq:dc1,eq:dc2}. The steps sequentially followed in the $i$th data-consistency block can be expressed as

\begin{align}
\label{eq:dc1}
    &x_{dc}^{i}    = \mathcal{F}^{-1}[\mathcal{F}[C\mathcal{U}(x_{p}^{i})] \odot (1-\mathcal{M}) +  \mathcal{F}[Cx_{us}] \odot \mathcal{M}]
    \\ &x_{dc, p}^{i} = \mathcal{P}(\text{SiLU}(x_{dc}^{i}))
\label{eq:dc2}
\end{align}
where $\mathcal{F}$ and $\mathcal{F}^{-1}$ denotes forward and backward 2D Fourier Transforms, respectively, $\mathcal{P}$ denotes the convolution-based patch embedding layer, $\mathcal{U}$ represents the unpatchify layer, $x_{p}^{i}$ is patchified input of the $i$th data-consistency layer, $x_{dc}^{i}$ is redundant intermediate output, $x_{dc, p}^{i}$ is patchified output of the $i$th data-consistency layer, and SiLU is the swish \cite{ramachandran2017searching} activation function.


In this design of architecture, we enable the information to be flown through patches in every possible directions, while keeping the reliance to the physical model by enforcing hard data-consistency between the VSSM blocks. 

\begin{figure*}[h]
    \centering
    \includegraphics[width=1\textwidth]{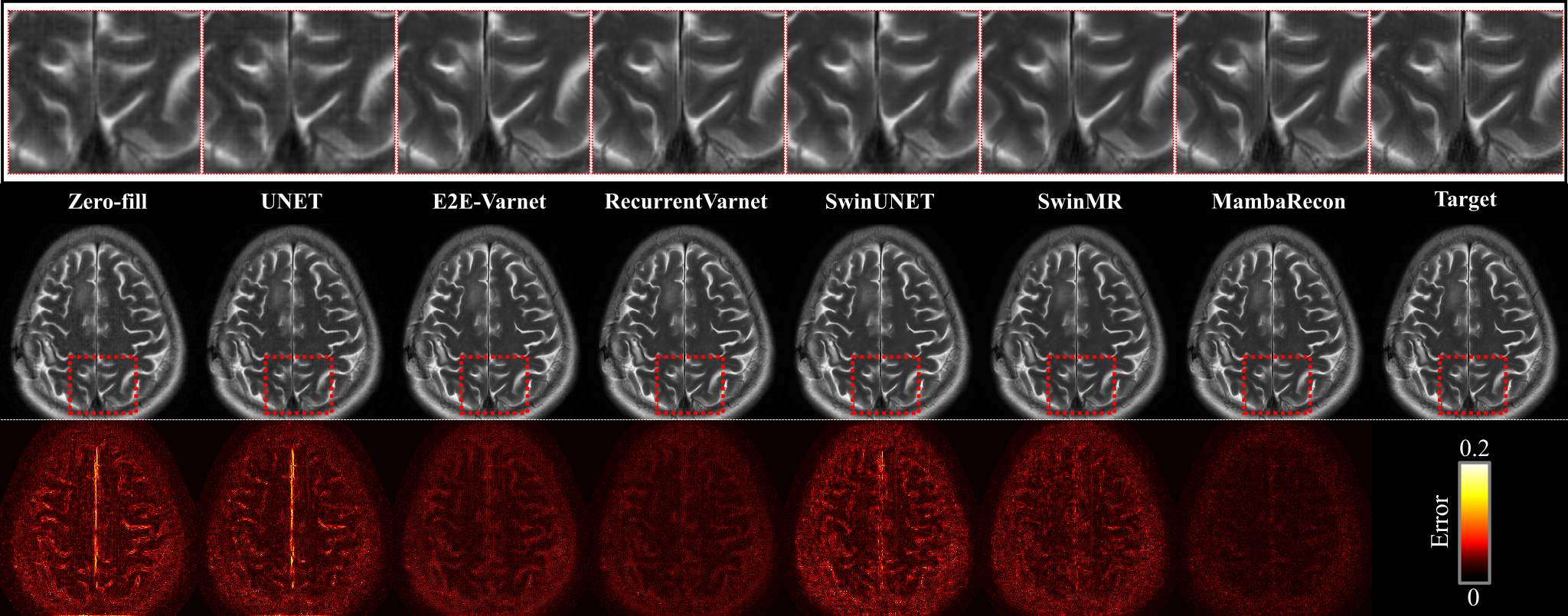}
    \caption{Reconstructions of T\SB{2} images with acceleration rate of 4 from fastMRI. Zoomed-in areas and error maps are attached on top and below of reconstructions.}
    \label{fig:fastmri_t2}
\end{figure*}

\begin{figure*}[h]
    \centering
    \includegraphics[width=1\textwidth]{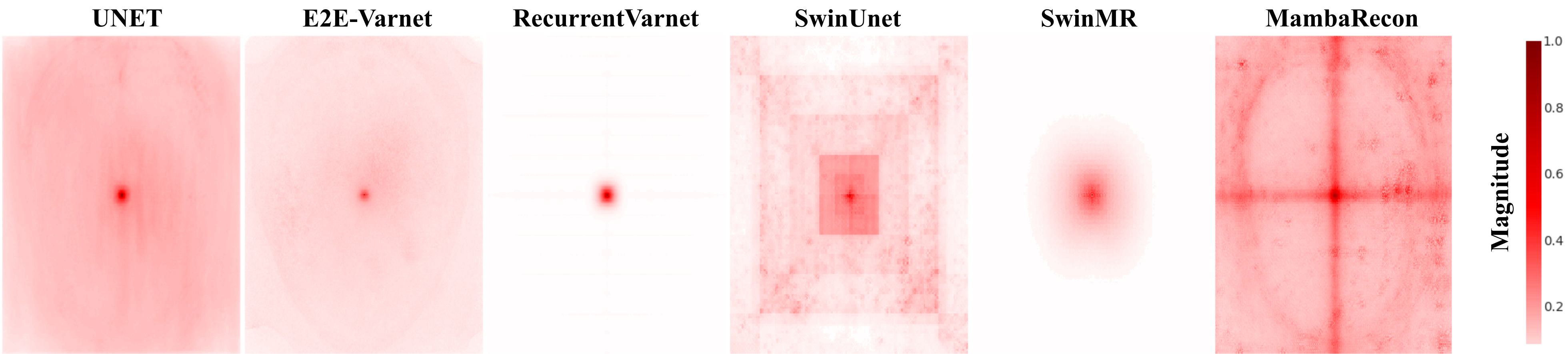}
    \caption{Effective receptive fields of each competing method are drawn after training, adapting the codes provided by \cite{liu2024vmamba}. Gradients are averaged across 100 test slices in fastMRI dataset. For physics-guided methods undersampling mask is given as zero to isolate the effect of the backbone. Each data point in these graphs represents the derivative of the center point in the reconstructed image with respect to the undersampled input image. A higher density of outer points indicates a stronger relationship between the center point and those locations, suggesting greater long-range sensitivity.}
    \label{fig:erf}
\end{figure*}

\subsection{Network and Training Details}
 AdamW optimizer is used for training with default parameters for 100,000 iterations with a batch size of 4. Learning rate is decayed after warm-up using half-cycle cosine decay starting from $1 \times 10^{-3}$ to $1 \times 10^{-6}$. Patch size is selected as 2 although 1, 4 and 8 are considered during hyper-parameter tuning as shown in ablation studies. Number of consecutive VSSM and data-consistency blocks are selected as 6; 4 and 8 are also considered as shown in the ablation studies. Inner state dimension of VSSM is selected as 16 following \cite{liu2024vmamba}, while hidden dimension of features is selected as 128. SiLU activation \cite{elfwing2018sigmoid} is used as the non-linearity throughout the network.

\begin{table*}[]
\caption{Reconstruction metrics shown in the IXI dataset when R = 4 and 8.}
\resizebox{\textwidth}{!}{%
\begin{tabular}{|c|cccccccccccc|}
\hline
                & \multicolumn{2}{c|}{T\SB{1}-4x}                                                & \multicolumn{2}{c|}{T\SB{1}-8x}                                                & \multicolumn{2}{c|}{T\SB{2}-4x}                                                & \multicolumn{2}{c|}{T\SB{2}-8x}                                                & \multicolumn{2}{c|}{PD-4x}                                                & \multicolumn{2}{c|}{PD-8x}                           \\ \hline
                & \multicolumn{12}{c|}{PSNR (dB) $\vert$ SSIM (\%)}                                                                                                                                                                                                                                                                                                                                                                                                      \\ \hline
UNET            & \multicolumn{1}{c|}{37.01}          & \multicolumn{1}{c|}{94.07}          & \multicolumn{1}{c|}{31.82}          & \multicolumn{1}{c|}{90.57}          & \multicolumn{1}{c|}{38.06}          & \multicolumn{1}{c|}{94.85}          & \multicolumn{1}{c|}{33.86}          & \multicolumn{1}{c|}{91.98}          & \multicolumn{1}{c|}{36.99}          & \multicolumn{1}{c|}{94.66}          & \multicolumn{1}{c|}{32.04}          & 91.43          \\ \hline
E2E-Varnet      & \multicolumn{1}{c|}{41.94}          & \multicolumn{1}{c|}{98.56}          & \multicolumn{1}{c|}{35.72}          & \multicolumn{1}{c|}{97.21}          & \multicolumn{1}{c|}{42.96}          & \multicolumn{1}{c|}{98.69}          & \multicolumn{1}{c|}{37.23}          & \multicolumn{1}{c|}{97.34}          & \multicolumn{1}{c|}{42.63}          & \multicolumn{1}{c|}{98.83}          & \multicolumn{1}{c|}{35.23}          & 97.59          \\ \hline
RecurrentVarnet & \multicolumn{1}{c|}{44.95}          & \multicolumn{1}{c|}{\textbf{99.48}} & \multicolumn{1}{c|}{37.78}          & \multicolumn{1}{c|}{98.55}          & \multicolumn{1}{c|}{46.24}          & \multicolumn{1}{c|}{\textbf{99.25}} & \multicolumn{1}{c|}{39.62}          & \multicolumn{1}{c|}{98.32}          & \multicolumn{1}{c|}{45.96}          & \multicolumn{1}{c|}{\textbf{99.34}} & \multicolumn{1}{c|}{37.96}          & 98.45          \\ \hline
SwinUNET        & \multicolumn{1}{c|}{34.77}          & \multicolumn{1}{c|}{94.73}          & \multicolumn{1}{c|}{30.59}          & \multicolumn{1}{c|}{91.75}          & \multicolumn{1}{c|}{37.54}          & \multicolumn{1}{c|}{95.11}          & \multicolumn{1}{c|}{33.92}          & \multicolumn{1}{c|}{92.44}          & \multicolumn{1}{c|}{32.58}          & \multicolumn{1}{c|}{94.71}          & \multicolumn{1}{c|}{29.32}          & 91.59          \\ \hline
SwinMR          & \multicolumn{1}{c|}{38.30}          & \multicolumn{1}{c|}{96.05}          & \multicolumn{1}{c|}{33.76}          & \multicolumn{1}{c|}{93.65}          & \multicolumn{1}{c|}{40.03}          & \multicolumn{1}{c|}{96.17}          & \multicolumn{1}{c|}{35.75}          & \multicolumn{1}{c|}{93.89}          & \multicolumn{1}{c|}{38.34}          & \multicolumn{1}{c|}{96.14}          & \multicolumn{1}{c|}{33.66}          & 93.61          \\ \hline
MambaRecon      & \multicolumn{1}{c|}{\textbf{45.59}} & \multicolumn{1}{c|}{98.52}          & \multicolumn{1}{c|}{\textbf{38.93}} & \multicolumn{1}{c|}{\textbf{98.75}} & \multicolumn{1}{c|}{\textbf{46.95}} & \multicolumn{1}{c|}{98.07}          & \multicolumn{1}{c|}{\textbf{40.75}} & \multicolumn{1}{c|}{\textbf{98.49}} & \multicolumn{1}{c|}{\textbf{46.81}} & \multicolumn{1}{c|}{98.28}          & \multicolumn{1}{c|}{\textbf{39.55}} & \textbf{98.63} \\ \hline
\end{tabular}%
}
\label{tab:ixi}
\end{table*}

\begin{figure*}[h]
    \centering
    \includegraphics[width=1\textwidth]{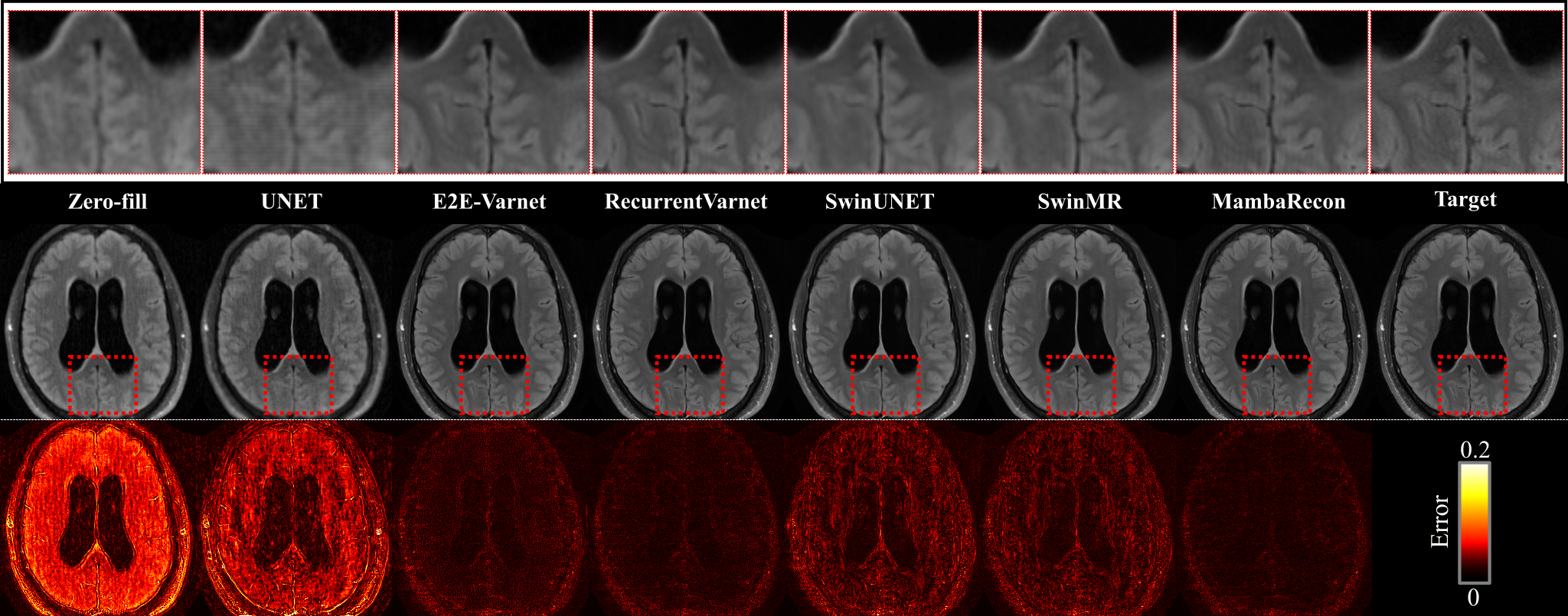}
    \caption{Reconstructions of Flair images with acceleration rate of 8 from fastMRI. Zoomed-in areas and error maps are attached on top and below of reconstructions.}
    \label{fig:fastmri_flair}
\end{figure*}

\section{Experiments}
Each competing method is trained end-to-end using under and fully-sampled image pairs. Acceleration is performed via variable-density 2D Gaussian masks with variance adjusted to achieve 4 and 8 acceleration rates. All experiments are performed on a single NVIDIA RTX A5000 gpu with a PyTorch framework. Number of parameters for each model has been shown in Table \ref{tab:param}.
\subsection{Datasets}
We used the following magnitude single-coil and complex multi-coil brain MRI datasets in our experiments. 
\begin{enumerate}
    \item \textbf{fastMRI}: Multi-coil brain MRI dataset \cite{fastmri} is considered, subjects are divided into training, validation and testing as (100, 10, 40). T\SB{1}-, T\SB{2}-weighted and Flair acquisitions are considered. There are no common protocol for all subjects. Number of coils are reduced to 5 to decrease the computational expense using \cite{zhang2013coil}. Coil sensitivity maps are estimated using ESPIRIT with default parameters \cite{uecker2014espirit}.
    \item \textbf{IXI}: Single-coil brain MRI data from IXI (http://brain-development.org/ixi-dataset/) is considered. T\SB{1}-, T\SB{2}- and PD-weighted acquisitions are used. 25 subjects are used for training, 5 for validation and 10 for testing in all experiments. 
\end{enumerate}

\begin{table}[]
\caption{The number of parameters for each competing model is presented, with MambaRecon having the fewest parameters.}
\centering
\resizebox{0.25\textwidth}{!}{%
\begin{tabular}{|c|c|}
\hline
Method & \#Parameters \\ \hline
UNET & 7.76e+6 \\ \hline
E2E-Varnet & 1.16e+7 \\ \hline
RecurrentVarnet & 2.83e+6 \\ \hline
SwinUNET & 2.71e+7 \\ \hline
SwinMR & 5.84e+6 \\ \hline
MambaRecon & 2.05e+6 \\ \hline
\end{tabular}%
}
\label{tab:param}
\end{table}

\subsection{Competing Methods}
\begin{itemize}
    \item \textbf{UNET}: U-shaped network model is utilized from \cite{fastmri}. Network hyper-parameters and implementation codes are gathered from \cite{DIRECTTOOLKIT}. Adam optimizer is used with the default parameters with a learning rate of 0.002, for 150,000 number of iterations.
    \item \textbf{E2E-Varnet}: End-to-end variational network is considered \cite{sriram2020end}. Network hyper-parameters and implementation codes are gathered from \cite{DIRECTTOOLKIT}. Adam optimizer is used with the default parameters with a learning rate of 0.0002, for 150,000 iterations.
    \item \textbf{RecurrentVarnet}: An unrolled k-space based recurrent variational model is considered \cite{yiasemis2022recurrent}. Network hyper-parameters and implementation codes are gathered from \cite{DIRECTTOOLKIT}. Adam optimizer is used with the default parameters with a learning rate of 0.0005, for 500,000 iterations.
    \item \textbf{SwinUnet}: A transformer based U-shaped segmentation model is adapted for MRI reconstruction \cite{cao2022swin}. AdamW optimizer is utilized for training with a learning rate of 0.0002 decayed after 5th epoch using cosine decay. Training is lasted for 100,000 iterations.
    \item \textbf{SwinMR}: A swin-transformer based MRI reconstruction model is considered \cite{huang2022swin}. Adam optimizer is used with the default parameters with a learning rate of 0.0002, for 100,000 iterations.
\end{itemize}

\begin{figure*}[h]
    \centering
    \includegraphics[width=1\textwidth]{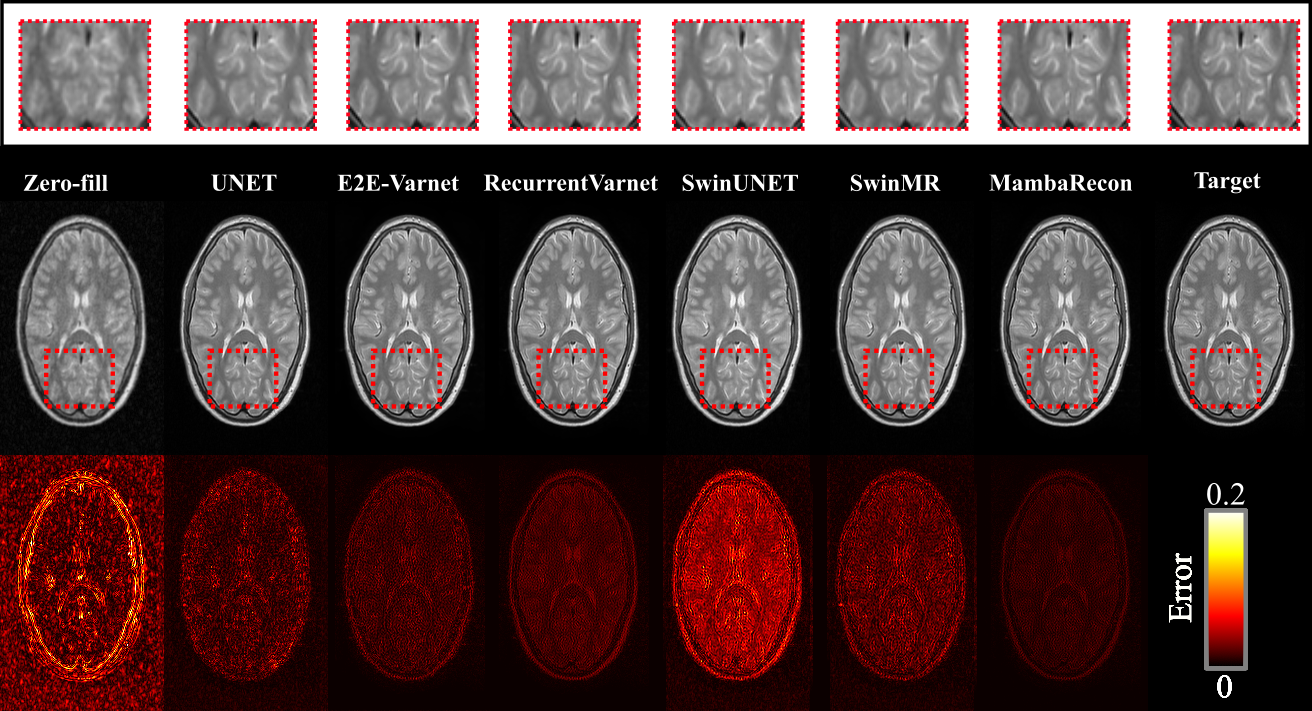}
    \caption{Reconstructions of PD images with acceleration rate of 4 from IXI. Zoomed-in areas and error maps are attached on top and below of reconstructions.}
    \label{fig:ixi_pd}
\end{figure*}

\begin{figure*}[h]
    \centering
    \includegraphics[width=1\textwidth]{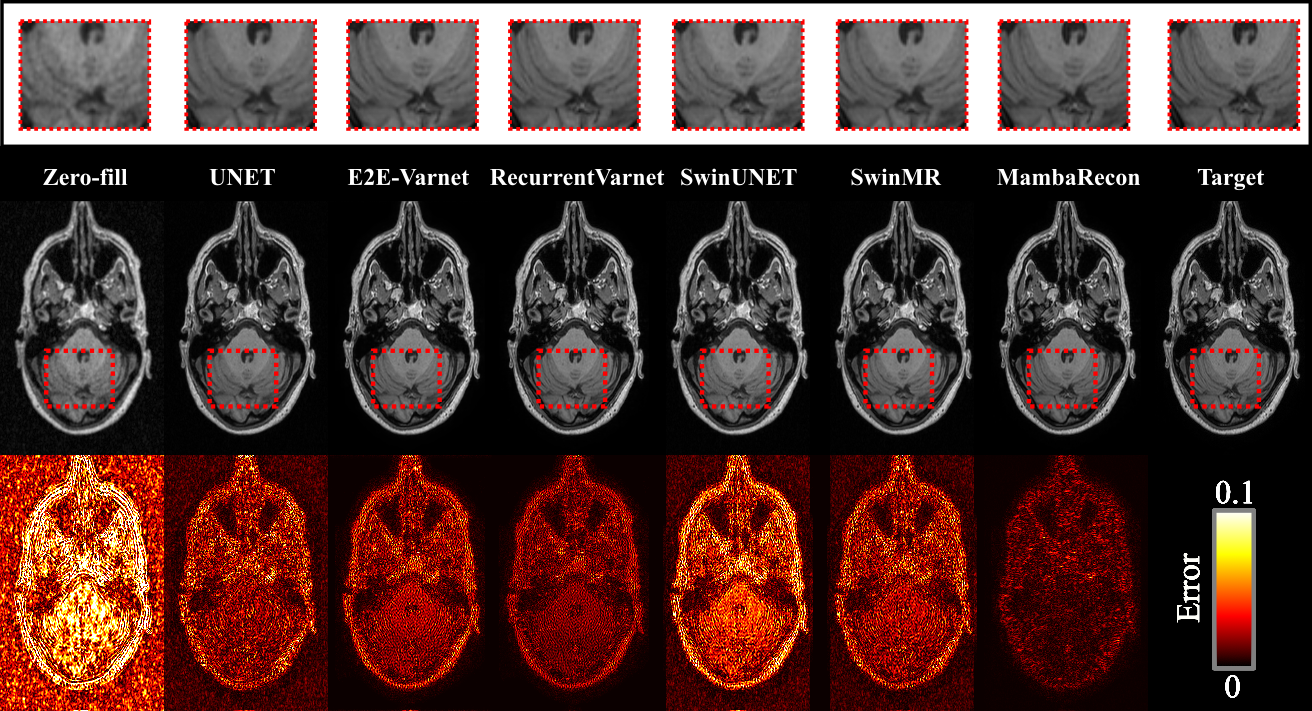}
    \caption{Reconstructions of T\SB{1}-weighted images with acceleration rate of 4 from IXI. Zoomed-in areas and error maps are attached on top and below of reconstructions.}
    \label{fig:ixi_t1}
\end{figure*}
\subsection{Ablation Studies}
\begin{table*}[]
\caption{Ablation results shown in fastMRI dataset when R = 4 and 8.}
\resizebox{\textwidth}{!}{%
\begin{tabular}{|c|cccccccccccc|}
\hline
           & \multicolumn{2}{c|}{T\SB{1}-4x}                                                & \multicolumn{2}{c|}{T\SB{1}-8x}                                                & \multicolumn{2}{c|}{T\SB{2}-4x}                                                & \multicolumn{2}{c|}{T\SB{2}-8x}                                                & \multicolumn{2}{c|}{FLAIR-4x}                                             & \multicolumn{2}{c|}{FLAIR-8x}                        \\ \hline
           & \multicolumn{12}{c|}{PSNR (dB) $\vert$ SSIM (\%)}                                                                                                                                                                                                                                                                                                                                                                                                      \\ \hline
OnlyDC     & \multicolumn{1}{c|}{39.10}          & \multicolumn{1}{c|}{96.25}          & \multicolumn{1}{c|}{34.53}          & \multicolumn{1}{c|}{92.93}          & \multicolumn{1}{c|}{34.65}          & \multicolumn{1}{c|}{94.49}          & \multicolumn{1}{c|}{31.52}          & \multicolumn{1}{c|}{90.94}          & \multicolumn{1}{c|}{35.44}          & \multicolumn{1}{c|}{94.13}          & \multicolumn{1}{c|}{30.95}          & 88.73          \\ \hline
SwinRecon  & \multicolumn{1}{c|}{43.53}          & \multicolumn{1}{c|}{97.68}          & \multicolumn{1}{c|}{38.40}          & \multicolumn{1}{c|}{95.30}          & \multicolumn{1}{c|}{38.94}          & \multicolumn{1}{c|}{97.19}          & \multicolumn{1}{c|}{34.86}          & \multicolumn{1}{c|}{94.81}          & \multicolumn{1}{c|}{38.79}          & \multicolumn{1}{c|}{96.19}          & \multicolumn{1}{c|}{34.34}          & 91.89          \\ \hline
$Depth=4$    & \multicolumn{1}{c|}{43.59}               & \multicolumn{1}{c|}{97.66}      & \multicolumn{1}{c|}{38.74}               & \multicolumn{1}{c|}{95.36}               & \multicolumn{1}{c|}{39.25}               & \multicolumn{1}{c|}{97.32}      & \multicolumn{1}{c|}{35.18}               & \multicolumn{1}{c|}{94.99}               & \multicolumn{1}{c|}{38.85}               & \multicolumn{1}{c|}{96.13}      & \multicolumn{1}{c|}{34.55}               &       91.90         \\ \hline
$Depth=8$    & \multicolumn{1}{c|}{43.61}          & \multicolumn{1}{c|}{97.74}          & \multicolumn{1}{c|}{39.00}          & \multicolumn{1}{c|}{\textbf{95.49}} & \multicolumn{1}{c|}{39.10}          & \multicolumn{1}{c|}{97.33}          & \multicolumn{1}{c|}{35.29}          & \multicolumn{1}{c|}{95.12}          & \multicolumn{1}{c|}{38.91}          & \multicolumn{1}{c|}{96.25}          & \multicolumn{1}{c|}{34.57}          & 92.19          \\ \hline
$p=1$        & \multicolumn{1}{c|}{43.67}          & \multicolumn{1}{c|}{\textbf{97.80}} & \multicolumn{1}{c|}{38.59}          & \multicolumn{1}{c|}{95.21}          & \multicolumn{1}{c|}{39.12}          & \multicolumn{1}{c|}{97.28}          & \multicolumn{1}{c|}{34.92}          & \multicolumn{1}{c|}{94.68}          & \multicolumn{1}{c|}{38.95}          & \multicolumn{1}{c|}{\textbf{96.36}} & \multicolumn{1}{c|}{34.38}          & 91.71          \\ \hline
$p=4$        & \multicolumn{1}{c|}{42.51}          & \multicolumn{1}{c|}{97.24}          & \multicolumn{1}{c|}{37.81}          & \multicolumn{1}{c|}{94.74}          & \multicolumn{1}{c|}{38.07}          & \multicolumn{1}{c|}{96.68}          & \multicolumn{1}{c|}{34.55}          & \multicolumn{1}{c|}{94.32}          & \multicolumn{1}{c|}{37.96}          & \multicolumn{1}{c|}{95.56}          & \multicolumn{1}{c|}{33.90}          & 91.27          \\ \hline
$p=8$        & \multicolumn{1}{c|}{39.83}          & \multicolumn{1}{c|}{96.22}          & \multicolumn{1}{c|}{36.39}          & \multicolumn{1}{c|}{93.86}          & \multicolumn{1}{c|}{35.61}          & \multicolumn{1}{c|}{95.15}          & \multicolumn{1}{c|}{33.29}          & \multicolumn{1}{c|}{92.96}          & \multicolumn{1}{c|}{35.80}          & \multicolumn{1}{c|}{93.78}          & \multicolumn{1}{c|}{32.73}          & 89.81          \\ \hline
MambaRecon & \multicolumn{1}{c|}{\textbf{43.93}} & \multicolumn{1}{c|}{97.63}          & \multicolumn{1}{c|}{\textbf{39.08}} & \multicolumn{1}{c|}{95.43}          & \multicolumn{1}{c|}{\textbf{39.43}} & \multicolumn{1}{c|}{\textbf{97.45}} & \multicolumn{1}{c|}{\textbf{35.47}} & \multicolumn{1}{c|}{\textbf{95.25}} & \multicolumn{1}{c|}{\textbf{39.21}} & \multicolumn{1}{c|}{\textbf{96.36}} & \multicolumn{1}{c|}{\textbf{34.84}} & \textbf{92.28} \\ \hline
\end{tabular}%
}
\label{tab:ablation}
\end{table*}
We perform a variety of ablation studies to show the individual effect of hyper-parameters and design choices. Performance metrics for ablation experiments are shown in Table \ref{tab:ablation}. As the metrics indicate, using an overly deep model or a very small patch size led to performance loss in addition to an increased computational budget. Here we list the ablation studies performed:
\begin{enumerate}
    \item OnlyDC: MambaRecon with only data-consistency blocks are considered where we removed the VSSM blocks completely from the model.
    \item SwinRecon: MambaRecon with swin transformer blocks is considered where we replaced VSSM blocks with swin transformer blocks with the same hidden dimension (128) and patch size ($p=2$). Window size is specified as 8.
    \item MambaRecon with more shallow and deeper backbones is considered with depth = 4 and 8 respectively. Here depth represents number of consecutive VSSM and data-consistency block pairs. 
    \item MambaRecon with different patch sizes is considered where $p$ ranged from 1 to 8.
\end{enumerate}

\section{Results}
We conducted a comprehensive comparison of MambaRecon against a diverse range of models and datasets to thoroughly evaluate its performance. Our benchmark included state-of-the-art physics-guided CNN-based models, E2E-Varnet and RecurrentVarnet, which leverage physical model for enhanced image reconstruction. We also included transformer-based reconstruction baselines, represented by SwinUNET and SwinMR, that utilize advanced transformer architectures to capture long range dependencies. Lastly, we considered a pure data-driven CNN-based network, exemplified by UNET, which relies solely on data for image reconstruction. We considered both multi-coil complex and single-coil magnitude MRI datasets represented by fastMRI and IXI respectively. This diverse selection allowed us to assess our model's capabilities across various reconstruction paradigms, from physics-guided approaches to advanced transformer-based and traditional CNN-based methods.

Peak Signal to Noise Ratio (PSNR) and Structural Similarity Index Measure (SSIM) \cite{wang2004image} are considered as the comparison metrics. Results are presented separately for each acceleration rate and contrast pair. Best results are highlighted as bold for each test case. Table \ref{tab:fastmri} and  Table \ref{tab:ixi} present the performance metrics for fastMRI and IXI respectively. Figure \ref{fig:fastmri_t2} shows representative reconstructions of a T\SB{2}-weighted slice with an acceleration rate of 4, Figure \ref{fig:fastmri_flair} stands for a Flair weighted slice when acceleration rate of 8 from fastMRI dataset. Figure \ref{fig:ixi_pd} and  Figure \ref{fig:ixi_t1} stand for PD and T\SB{1}-weighted reconstructions when acceleration rate of 4 from IXI dataset. 

Quantitatively, compared with the second best method; MambaRecon yields 0.72dB more average PSNR in the fastMRI dataset and 1.01dB more average PSNR in the IXI dataset. Visually, MambaRecon captured almost all of the high frequency details hence yields the darkest error maps in the reconstruction figures without suffering from over-smoothing effect or noise. We also illustrate the effective receptive field of each competing method in Figure \ref{fig:erf}. We observe the most powerful receptive field corresponds to MambaRecon which yields a global receptive field spanning the whole image. We present the number of parameters in each competing method in Table \ref{tab:param}.


\section{Discussion and Future Work}

We proposed a lightweight and efficient reconstruction model combining SSMs with phyics-guided MRI models. To overcome the limitations of the original SSMs' single-direction spanning, we adopted a multi-directional spanning approach. However, this approach might be less optimal than self-attention transformers in fully capturing the interactions between image patches. Implementing a more advanced scanning technique could enhance the model's performance. 

Our method currently relies on ESPIRIT \cite{uecker2014espirit} for coil sensitivity estimation. By integrating this process within a unified framework, we could enable end-to-end training, potentially enhancing the efficiency and accuracy of the overall model. This unified approach would streamline the workflow, allowing for simultaneous optimization of both the coil sensitivity estimation and the subsequent image reconstruction, thereby improving performance. 

Our model, like all other competing methods, is trained in a supervised manner using under-sampled and fully-sampled images from public MRI datasets. To eliminate the dependency on supervised training, a self-supervised training loss, as demonstrated by \cite{yaman2020}, could be employed. Moreover, a test-time adaptation approach could be beneficial to incorporate the changes in physical-model like acceleration rate or under-sampling pattern. We leave this exploration for future work.

\section{Conclusion}
We proposed a physics-guided MRI reconstruction framework utilizing structured state space models in its core. Our model benefits from high long range contextual sensitivity without creating a computational burden and yields a better reconstruction quality compared with the state-of-the-art reconstruction baselines.

\newpage

{\small
\bibliographystyle{ieee_fullname}
\bibliography{main}
}

\end{document}